\documentclass[aps,prd,showpacs,twocolumn,groupedaddress]{revtex4}
\usepackage{graphicx}

\begin{document}

\title{Lattice study of the chiral magnetic effect in a chirally imbalanced matter}

\author{Arata~Yamamoto}
\affiliation{Department of Physics, The University of Tokyo, Tokyo 113-0033, Japan}

\date{\today}

\begin{abstract}
We investigate the chiral magnetic effect by lattice QCD with a chiral chemical potential.
In a chirally imbalanced matter, we obtain a finite induced current along an external magnetic field.
We analyze the dependence on the lattice spacing, the temperature, the spatial volume, and the fermion mass.
The present result indicates that the continuum limit is important for the quantitative argument of the strength of the induced current.
\end{abstract}


\maketitle

\section{Introduction}
Topology plays a significant role in classical and quantum field theory.
In quantum chromodynamics (QCD), the nontrivial topology of the background gauge field is related to the axial anomaly of the fermion.
From the Atiyah-Singer index theorem,
\begin{eqnarray}
N_f Q = N_R - N_L
\end{eqnarray}
\cite{Atiyah:1968mp}.
In the QCD vacuum, the total numbers of the left-handed and right-handed fermions are the same, and the average topological charge is zero.
However, the topological charge strongly fluctuates in space-time.
This topological fluctuation is essential for the $\eta'$-meson mass \cite{Witten:1979vv,Veneziano:1979ec}.

The chiral magnetic effect is one possible candidate to detect the topological fluctuation in experiments \cite{Kharzeev:2004ey,Kharzeev:2007tn,Kharzeev:2007jp,:2009uh}.
A noncentral heavy-ion collision produces a very strong magnetic field.
The fermion flows along this magnetic field, and its direction is determined from its chirality.
The topological fluctuation generates a chiral imbalance in some local domain, and then it induces a charged flow of the fermion, or equivalently, an electric current.
When all experimental events are averaged over, the total electric current is zero, but its event-by-event correlation is nonzero.

Several theoretical works have studied the chiral magnetic effect in lattice QCD \cite{Buividovich:2009zzb,Buividovich:2009wi,Abramczyk:2009gb,Braguta:2010ej,Ishikawa:2011}.
These works have tried to measure current-current correlation or charge density distribution.
In order to analyze the chiral magnetic effect in this direction, it is necessary to take into account topological objects on the lattice, however, it is usually difficult in practice.
We overcome the difficulty by introducing chiral chemical potential, instead of the topological fluctuation.

In this paper, we investigate the chiral magnetic effect in lattice QCD by extending a previous work \cite{Yamamoto:2011gk}.
We discuss a theoretical background of the chiral chemical potential in Sec.~II, and the lattice QCD formalism in Sec.~III.
We show the numerical result of the full QCD simulation in Sec.~IV and of the quenched QCD simulation in Sec.~V.
Finally, we conclude in Sec.~VI.

\section{Chiral chemical potential}
The chiral chemical potential $\mu_5$ is defined in the continuum Dirac operator as
\begin{eqnarray}
D(\mu_5) = \gamma_\mu (\partial_\mu +ig t^a A^a_\mu) + m + \mu_5 \gamma_4 \gamma_5
\label{eqDiracC}
\end{eqnarray}
\cite{Fukushima:2008xe}.
The Euclidean metric is used throughout this paper.
The chiral chemical potential couples to the chiral charge density
\begin{eqnarray}
n_5 &\equiv& \frac{T}{V}\frac{\partial}{\partial \mu_5} \ln Z
= - \frac{1}{V} \int d^3x \langle \bar{\psi} \gamma_4 \gamma_5 \psi \rangle \nonumber \\
&=& \frac{1}{V} \int d^3x \langle \psi^\dagger_R \psi_R - \psi^\dagger_L \psi_L \rangle
= \frac{1}{V} (N_R - N_L) \ .
\end{eqnarray}
The chiral charge is the number difference between the right-handed and left-handed fermions.
Strictly speaking, the chiral charge is not a conserved quantity in QCD because of the axial anomaly.
The chiral chemical potential realizes a finite chiral charge in an equilibrium state, namely, a chirally imbalanced matter.
This is useful for theoretical treatment, for example, the lattice QCD simulation is possible.

The chiral chemical potential has been introduced for analyzing the chiral magnetic effect \cite{Fukushima:2008xe,Kharzeev:2009pj,Fukushima:2010fe,Fukushima:2010zza}.
The induced vector current was derived,
\begin{eqnarray}
J = \frac{1}{2\pi^2} \mu_5 qB \ ,
\label{eqj}
\end{eqnarray}
from the one-component Dirac equation coupled with the background magnetic field \cite{Fukushima:2008xe}.
(Note that the overall factor is different from its original expression by $q$.
This is a matter of definition.
The electric current is usually defined as $J_{\rm EM} = q J$.
)
This formula is independent of the fermion mass and the temperature.

The significant property of the chiral chemical potential is that the sign problem does not arise.
In lattice QCD, a quark chemical potential causes the sign problem.
The naive Monte Carlo algorithm breaks down at a finite quark number density.
Although many numerical methods have been proposed for simulating the quark chemical potential, their applicabilities are limited within a small quark chemical potential \cite{Stephanov:2007fk}.
For a large chemical potential, we can only simulate exceptional cases, such as the two-color QCD and an isospin chemical potential, which can exactly avoid the sign problem \cite{Hands:1999md,Kogut:2001na,Kogut:2002tm,Nakamura:2003gj,deForcrand:2007uz}.
The chiral chemical potential is an interesting possibility to study the finite density matter in lattice QCD.

\section{LATTICE QCD FORMALISM}
For the numerical simulation, we used the SU(3) plaquette gauge action and the Wilson fermion action.
We considered the degenerate two-flavor case, in which the two fermions have the same mass $m$ and charge $q$.
This approximation simplifies the simulation algorithm, especially, the hybrid Monte Carlo algorithm.

The lattice Dirac operator of the Wilson fermion is
\begin{eqnarray}
D_{\rm W}(\mu_5) &=& 1
- \kappa \sum_i \bigl[ (1-\gamma_i)T_{i+} \nonumber + (1+\gamma_i)T_{i-} \bigr] \nonumber\\
&&- \kappa \bigl[ (1-\gamma_4 e^{a\mu_5\gamma_5})T_{4+} \nonumber\\
&&+ (1+\gamma_4 e^{-a\mu_5\gamma_5})T_{4-} \bigr] 
\label{eqDirac}
\end{eqnarray}
with
\begin{eqnarray}
&&[T_{\mu +}]_{x,y} \equiv U_\mu (x) \delta_{x+\hat{\mu},y}
\label{eqpp} \\
&&[T_{\mu -}]_{x,y} \equiv U^\dagger_\mu (y) \delta_{x-\hat{\mu},y}
\label{eqpm} \\
&&e^{\pm a\mu_5\gamma_5} = \cosh (a\mu_5) \pm \gamma_5\sinh (a\mu_5) \ .
\end{eqnarray}
This Dirac operator reproduces the continuum form (\ref{eqDiracC}), apart from an overall factor, in the naive continuum limit $a\to 0$.
This Dirac operator satisfies the ``$\gamma_5$-Hermite'' property
\begin{eqnarray}
\gamma_5 D(\mu_5) = [ \gamma_5 D (\mu_5) ]^\dagger \ ,
\label{eqher}
\end{eqnarray}
or equivalently,
\begin{eqnarray}
\gamma_5 D(\mu_5) \gamma_5 =  D^\dagger (\mu_5)\ .
\end{eqnarray}
From this property, the fermion determinant is real and semi-positive in the two-flavor case,
\begin{equation}
\det \left(
\begin{array}{cc}
D(\mu_5) & 0 \\
0 & D(\mu_5)
\end{array}
  \right) = |\det D(\mu_5)|^2 \ge 0 \ .
\end{equation}
Thus, the sign problem does not occur.

To apply an external magnetic field, we also need the U(1) gauge field.
On the lattice, the U(1) gauge field is given as the Abelian phase factor $u_\mu(x) = \exp(iaqA_\mu(x))$.
Since the magnetic field is external, the field strength term of the U(1) gauge field is not introduced.
The U(1) gauge field is introduced only in the Dirac operator (\ref{eqDirac}) by replacing
\begin{eqnarray}
U_\mu(x) \to u_\mu(x) U_\mu(x) \ .
\end{eqnarray}
In a finite-volume lattice with periodic boundary conditions, the quantized value of the magnetic field is allowed as
\begin{eqnarray}
a^2qB=\frac{2\pi}{N_s^2} \times {\rm (integer)} \ .
\end{eqnarray}
\cite{AlHashimi:2008hr}.
For a homogeneous magnetic field in the $x_3$-direction, the phase factors are
\begin{eqnarray}
 u_1(x)&=&\exp(-iaqBN_sx_2) \quad {\rm at}\ x_1=aN_s  \\
 u_2(x)&=&\exp(iaqBx_1) \\
 u_\mu(x)&=&1 \quad {\rm for}\ {\rm other}\ {\rm components}\ .
\end{eqnarray}
The magnetic field $B$ and the charge $q$ do not appear separately.
The combination $a^2qB$ is an input parameter of the simulation.

\section{Full QCD analysis}

\begin{table}[t]
\renewcommand{\tabcolsep}{0.4pc} 
\renewcommand{\arraystretch}{1} 
\caption{Simulation parameters of full QCD \cite{Orth:2005kq}.}
\label{tab1}
\begin{tabular}{cccccc}
\hline\hline
$\beta$ & $a^{-1}$ & $\kappa$ & $m_{\rm PS}/m_{\rm V}$ & $N_s$ & $N_t$ \\
\hline
5.32144 & 1.5 GeV & 0.16650 & 0.5 & 12 & 4\\
\hline\hline
\end{tabular}
\end{table}

\begin{figure}[b]
\begin{center}
\includegraphics[scale=1.2]{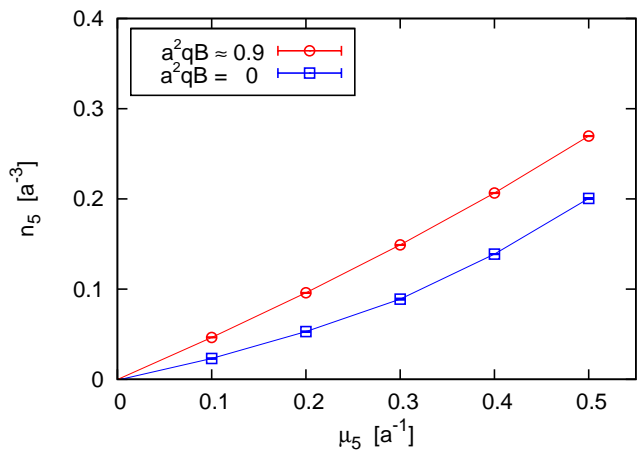}
\caption{The chiral charge density $n_5$ in full QCD.}
\label{fig1}
\end{center}
\end{figure}

\begin{figure}[t]
\begin{center}
\includegraphics[scale=1.2]{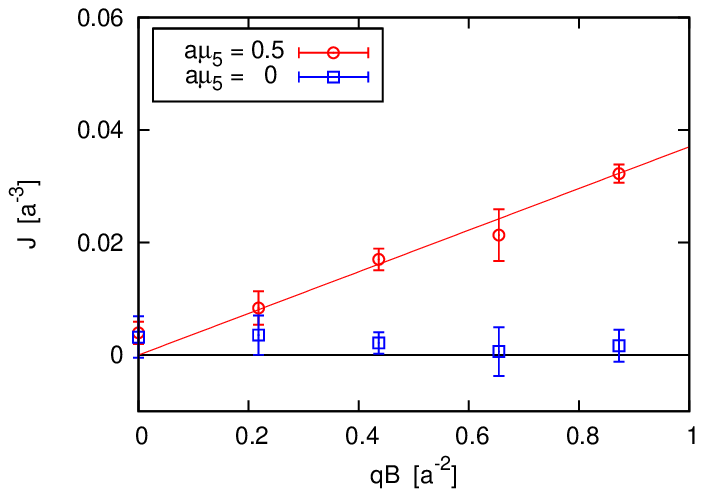}
\caption{The vector current density $J$ as a function of the magnetic field $qB$ in full QCD.}
\label{fig2}
\end{center}
\begin{center}
\includegraphics[scale=1.2]{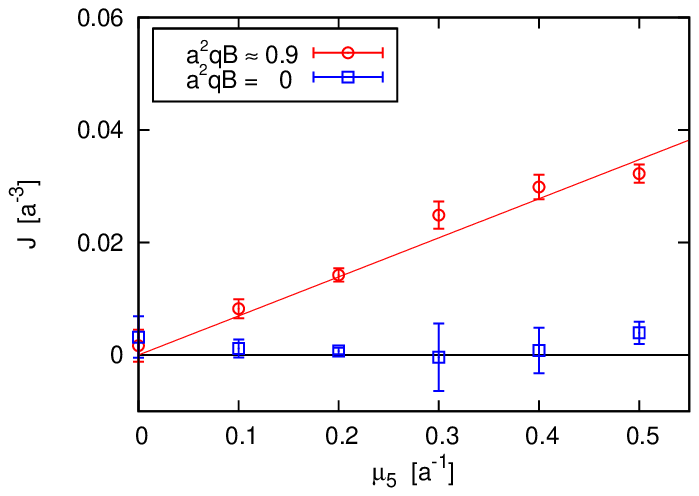}
\caption{The vector current density $J$ as a function of the chiral chemical potential $\mu_5$ in full QCD.}
\label{fig3}
\end{center}
\end{figure}

In this section, we show the numerical results in full QCD.
To generate the dynamical gauge ensembles, we used the hybrid Monte Carlo algorithm.
The lattice spacing $a^{-1} \simeq 1.5$ GeV ($a \simeq 0.13$ fm) and the mass ratio of pion to $\rho$-meson is $m_{\rm PS}/m_{\rm V} \simeq 0.5$ \cite{Orth:2005kq}.
The simulation parameters are listed in Table \ref{tab1}.
The physical temperature is $T = 1/(N_ta) \simeq 400$ MeV, which is in the deconfinement phase.

In Fig.~\ref{fig1}, we plot the chiral charge density
\begin{eqnarray}
n_5 = - \frac{1}{V} \sum_{\rm site} \left \langle \bar{\psi} \gamma_4\gamma_5 \psi \right \rangle \ .
\label{eqn5}
\end{eqnarray}
At a very large chiral chemical potential, the so-called saturation, which is a lattice artifact, appears \cite{Yamamoto:2011gk}.
To avoid this problem, we restricted the chiral chemical potential in $a\mu_5 \le 0.5$.
A finite chiral chemical potential generates a finite chiral charge density.
The chiral charge density is enhanced when the magnetic field is applied.
This is because the thermodynamic potential is increased by the magnetic field \cite{Fukushima:2008xe}.

We calculated the local vector current density
\begin{eqnarray}
J =  \frac{1}{V} \sum_{\rm site} \langle \bar{\psi} \gamma_3 \psi \rangle
\label{eqlvc}
\end{eqnarray}
along the magnetic field.
The transverse components are always zero, $\langle \bar{\psi} \gamma_1 \psi \rangle = \langle \bar{\psi} \gamma_2 \psi \rangle = 0$, because they are irrelevant for the chiral magnetic effect \cite{Yamamoto:2011gk}.
The vector current density is plotted as a function of the magnetic field in Fig.~\ref{fig2} and of the chiral chemical potential in Fig.~\ref{fig3}.
The vector current density is a linearly increasing function both of the magnetic field and of the chiral chemical potential.
We parametrize the induced current density as
\begin{eqnarray}
J = N_{\rm dof} C \mu_5 qB \ .
\label{eqjlat}
\end{eqnarray}
The factor $N_{\rm dof} = N_c \times N_f =6$ is the number of fermions with the same charge.
This functional form is consistent with the analytic formula (\ref{eqj}).
In the analytic formula, the overall coefficient is $1/(2\pi^2) \simeq 0.05$.
On the other hand, the best-fit value of the lattice data is $C = 0.013 \pm 0.001$.
The induced current seems to be suppressed.
However, in order to compare the overall coefficients quantitatively, we have to estimate several systematic effects, e.g., the renormalization.
In the next section, we analyze such systematic effects in quenched QCD.

\section{Quenched QCD analysis}

\begin{table}[t]
\renewcommand{\tabcolsep}{0.4pc} 
\renewcommand{\arraystretch}{1} 
\caption{Simulation parameters of quenched QCD \cite{Aoki:1999yr}.}
\label{tab2}
\begin{tabular}{cccccc}
\hline\hline
$\beta$ & $a^{-1}$ & $\kappa$ & $m_{\rm PS}/m_{\rm V}$ & $N_s$ & $N_t$ \\
\hline
5.90 & 1.9 GeV & 0.15920 & 0.4 & 12, 16, 20 & 4\\
5.90 & 1.9 GeV & 0.15890 & 0.5 & 12 & 4\\
5.90 & 1.9 GeV & 0.15740 & 0.7 & 12 & 4, 6, 12\\
6.25 & 3.1 GeV & 0.15115 & 0.7 & 18 & 6, 10, 18\\
6.47 & 4.0 GeV & 0.14885 & 0.7 & 24 & 8\\
\hline\hline
\end{tabular}
\end{table}

\begin{figure}[b]
\begin{center}
\includegraphics[scale=1.2]{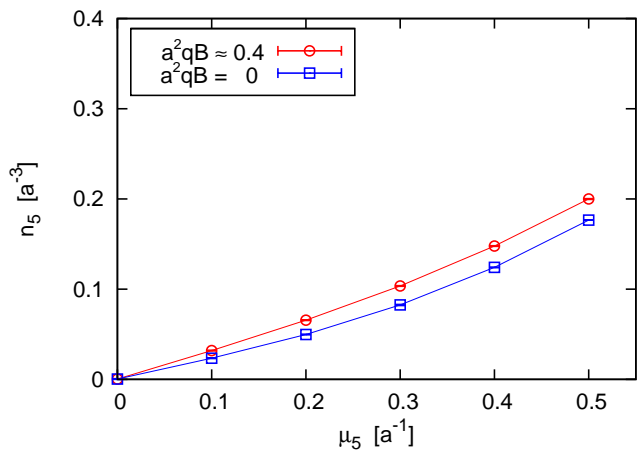}
\caption{The chiral charge density $n_5$ in quenched QCD.
The physical temperature is about $T \simeq 500$ MeV.
The meson mass ratio is $m_{\rm PS}/m_{\rm V} \simeq 0.4$.}
\label{fig4}
\end{center}
\end{figure}

We performed the quenched QCD simulation.
The quenched approximation is to neglect the fermion determinant in the Monte Carlo sampling.
In quenched QCD, the computational cost is highly reduced and the parameter tuning is very easy.
This approximation is considerably reasonable in many cases, e.g., hadron spectrum.
We expect the quenched approximation to work at least for a small chiral chemical potential, and evaluate systematic effects on the induced current.
The used gauge configurations are summarized in Table \ref{tab2} \cite{Aoki:1999yr}.

\begin{figure}[t]
\begin{center}
\includegraphics[scale=1.2]{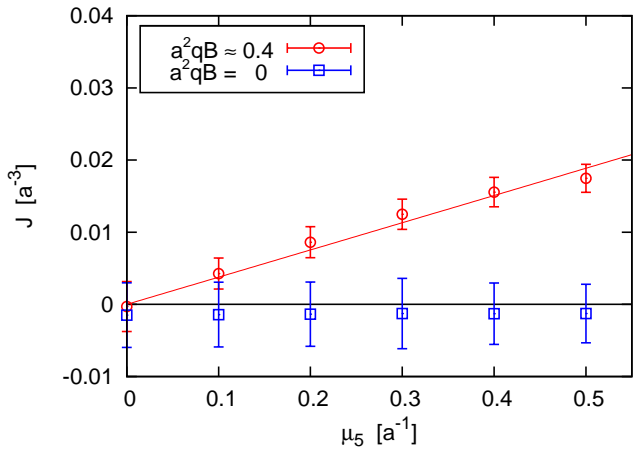}
\caption{The vector current density $J$ in quenched QCD.
The simulation parameters are the same as in Fig.~\ref{fig4}}
\label{fig5}
\end{center}
\end{figure}

We show the chiral charge density in Fig.~\ref{fig4} and the vector current density in Fig.~\ref{fig5}.
The qualitative behavior is consistent with the full QCD simulation.
The induced current shows the linearly rising behavior.
We fit the induced current by Eq.~(\ref{eqjlat}) and analyze the parameter dependence of the overall coefficient $C$.

First, we focus on the problem of the discretization.
The local vector current is not renormalization-group invariant on the lattice.
This is different from the continuum theory, in which the local vector current is strictly renormalization-group invariant.
The conserved vector current on the lattice is a point-split-form Green function \cite{Karsten:1980wd}.
To obtain the renormalization-group invariant form, we have to calculate the renormalization factor \cite{Maiani:1986yj,Martinelli:1994ty}.
In general, a large statistics is needed to calculate the renormalization factor of the flavor-singlet vector current because the disconnected contribution is rather noisy.
Instead of this, we analyze the dependence of the overall coefficient on the lattice spacing $a$.
In the continuum limit $a \to 0$, the renormalization factor automatically approaches to 1, and thus the ambiguity of the renormalization factor disappears.
The lattice spacing dependence is also important to estimate other lattice discretization artifacts.
For example, the $O(a)$ term of the Wilson fermion action explicitly breaks the chiral symmetry.
Such lattice artifacts also disappear in the continuum limit.

\begin{figure}[t]
\begin{center}
\includegraphics[scale=1.2]{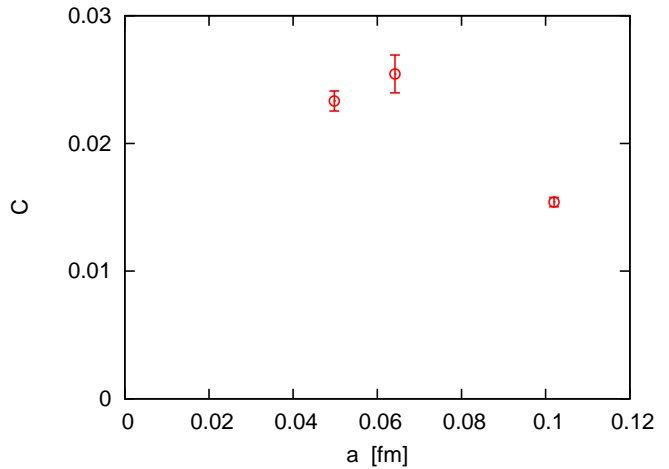}
\caption{Lattice spacing $a$ dependence.
The physical temperature is about $T \simeq 500$ MeV.
The meson mass ratio is $m_{\rm PS}/m_{\rm V} \simeq 0.7$.}
\label{fig6}
\end{center}
\end{figure}

We calculated the induced current at $\beta = 5.90$, 6.25, and 6.47.
By tuning the simulation parameters, we fixed the spatial volume, the temperature, and the meson masses. 
In Fig.~\ref{fig6}, we plot the best-fit value of the overall coefficient $C$.
The overall coefficient depends on the lattice spacing.
Its value is close to the full QCD result in $a > 0.1$ fm and larger in $a < 0.1$ fm.
This result indicates that it is important to reduce the discretization effect and to take the continuum extrapolation for more quantitative argument.
Although we cannot determine the functional form of the extrapolation, the overall coefficient seems $C \simeq 0.02$ - 0.03 in the continuum limit.

\begin{figure}[t]
\begin{center}
\includegraphics[scale=1.2]{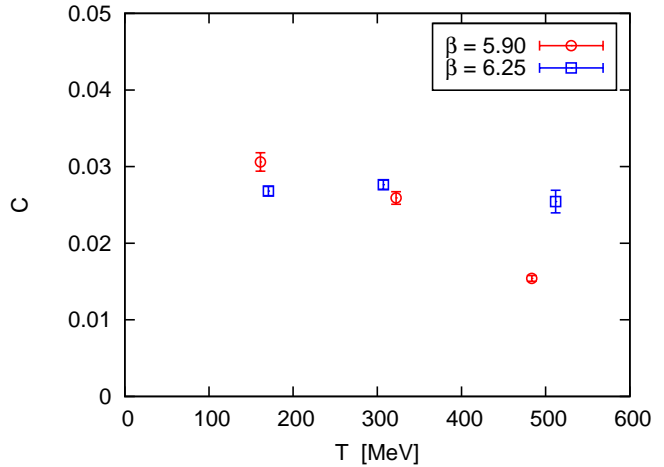}
\caption{Temperature $T$ dependence.
The phase transition temperature is about 270 MeV.
The meson mass ratio is $m_{\rm PS}/m_{\rm V} \simeq 0.7$.}
\label{fig7}
\end{center}
\end{figure}

In Fig.~\ref{fig7}, we plot the data of $\beta = 5.90$ and 6.25 as a function of the temperature $T=1/(N_ta)$.
Although the unrenormalized values need not be the same between the different lattice spacings, the qualitative behaviors should be consistent.
The data of $\beta = 5.90$ and $N_t=4$ ($T\simeq 500$ MeV) deviates from the others.
The reason would be that the discretization artifact of the unimproved fermion is sizable at $N_t=4$, which is known in the standard lattice QCD at finite temperature.
Except for $N_t=4$, the induced current is not so sensitive to the temperature.
The induced current is nonzero even below the phase transition temperature $T_c \simeq 270$ MeV, i.e., in the confinement phase.
Basically, the chiral magnetic effect is expected in the deconfinement phase because colored particles cannot flow independently in the confinement phase.
Such a real-time information is, however, not reflected in the local value of the induced current in an equilibrium state.

\begin{figure}[t]
\begin{center}
\includegraphics[scale=1.2]{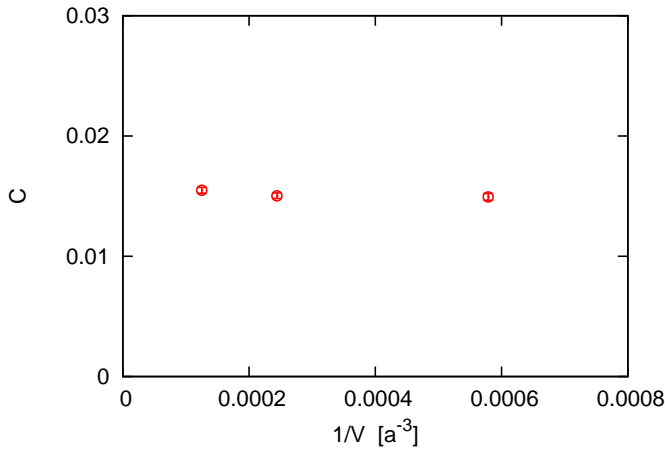}
\caption{Volume $V$ dependence.
The physical temperature is about $T \simeq 500$ MeV.
The meson mass ratio is $m_{\rm PS}/m_{\rm V} \simeq 0.4$.}
\label{fig8}
\end{center}
\end{figure}

Next, we show the volume dependence in Fig.~\ref{fig8}.
The lattice simulation is performed in a finite box, not in the infinite volume.
The spatial volume should be large enough that the obtained result is independent of the spatial volume.
In order to change the spatial volume $V=a^3N_s^3$, we fixed the lattice spacing $a$ and changed the spatial extent as $N_s = 12$, 16, and 20.
As expected, the induced current is independent of the spatial volume.

\begin{figure}[t]
\begin{center}
\includegraphics[scale=1.2]{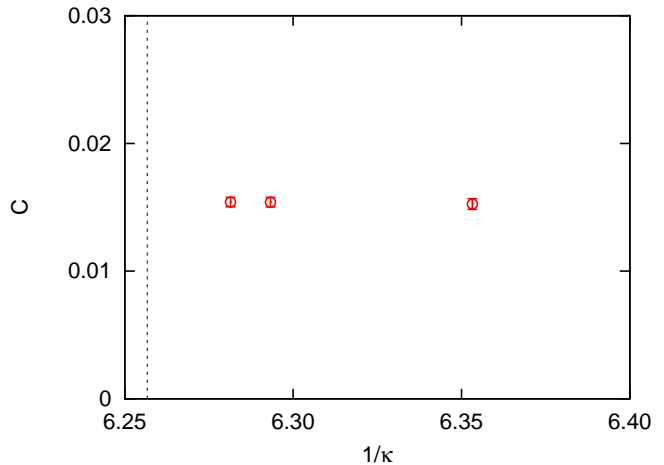}
\caption{Hopping parameter $\kappa$ dependence.
The dotted line corresponds to the critical hopping parameter $\kappa_c=0.15983$.
The physical temperature is about $T \simeq 500$ MeV.}
\label{fig9}
\end{center}
\end{figure}

Finally, we analyze the dependence on the hopping parameter, which is equivalent to the fermion mass dependence.
The result is shown in Fig.~\ref{fig9}.
The used hopping parameters are $\kappa = 0.15920$, 0.15890, and 0.15740, which correspond to $m_{\rm PS}/m_{\rm V}=0.4$, 0.5, and 0.7, respectively. 
The critical hopping parameter is $\kappa_c=0.15983$, which corresponds the chiral limit \cite{Aoki:1999yr}.
The induced current is almost independent of the fermion mass.
This is consistent with the analytic formula (\ref{eqj}).

\section{Summary and outlook}

We have analyzed the chiral magnetic effect by the lattice QCD simulation with the chiral chemical potential.
At the qualitative level, the induced current is consistent with the analytic formula which is derived from the Dirac equation coupled with an external magnetic field.
At the quantitative level, the induced current is somehow suppressed compared to the analytic formula.
The estimated value of the overall coefficient is $C \simeq 0.02$ - 0.03 even after the continuum extrapolation.
Although the systematic effects have been estimated in quenched QCD, the situation will be more or less the same in full QCD.
In order to reduce the discretization effect, we should first introduce the improved fermion action, since the fermion action has a larger discretization error than the gauge action.
More quantitative calculation will shed light on possible QCD corrections to the chiral magnetic effect \cite{Fukushima:2010zza}.

We have not discussed about chiral symmetry.
Since the Wilson fermion explicitly breaks chiral symmetry at a finite lattice spacing, we cannot analyze chiral symmetry in the present calculation.
We show the functional forms of other lattice Dirac operators in Appendix A.
By using lattice Dirac operators with better chiral property, we can study the role of chiral symmetry for the chiral magnetic effect.

\section*{ACKNOWLEDGMENTS}
The author would like to acknowledge K.~Fukushima, D.~E.~Kharzeev, T.~Hatsuda, and S.~Sasaki.
This work was supported in part by the Grant-in-Aid for Scientific Research in Japan under Grant No.~22340052.
The lattice QCD simulations were carried out on NEC SX-8R in Osaka University.

\appendix
\section{Lattice Dirac operators}
The simplest choice of the lattice Dirac operator with the chiral chemical potential is the Wilson Dirac operator (\ref{eqDirac}).
The chiral chemical potential can be implemented for other lattice fermions with better chirality, e.g., the staggered fermion, the domain-wall fermion and the overlap fermion.

In the staggered fermion, the Dirac spinor is constructed by mixing spinorless Grassmann fields on different lattice sites.
For example, in four dimensions, a four-taste four-component Dirac spinor is constructed from $2^4$ lattice sites.
The gamma matrix appears as a direct product $(\gamma_a \otimes \gamma_b)$ of two matrices which act on the spinor space and the taste space, respectively.
A naive choice of the staggered Dirac operator is
\begin{eqnarray}
D_{\rm KS}(\mu_5) &=& ma + \frac{1}{2} \sum_i \eta_i ( T_{i+} - T_{i-} ) \nonumber\\
&&+ \frac{1}{2} \eta_4 ( T_{4+} e^{a\mu_5\Gamma_5} - T_{4-} e^{-a\mu_5\Gamma_5} ) \ ,
\end{eqnarray}
with the staggered phase factor
\begin{eqnarray}
[\eta_\mu]_{x,y} = (-1)^{x_1 + \cdots + x_{\mu-1}} \delta_{x,y} \ .
\end{eqnarray} 

We consider two types of $\Gamma_5$.
To construct the taste-singlet matrix $(\gamma_5 \otimes 1)$, we take
\begin{eqnarray}
\Gamma_5 = \left[ \prod_{\mu} \eta_\mu \frac{T_{\mu+}+T_{\mu-}}{2} \right]_{\rm sym},
\end{eqnarray}
where ``sym'' means the symmetric average of the path-ordered products.
The matrix factor $e^{\pm a\mu_5\Gamma_5}$ is a product of hopping terms, and thus nonlocal.
This is understood from the fact that the operation of the gamma matrix corresponds to mixing different lattice sites in the staggered fermion.
Such a nonlocal Dirac operator is expensive for the numerical simulation, especially for the dynamical simulation.
For a local realization, we take
\begin{eqnarray}
[\Gamma_5]_{x,y} =[\eta_5]_{x,y} \equiv (-1)^{x_1 + x_2 + x_3 + x_4} \delta_{x,y} \ .
\end{eqnarray}
This matrix $\Gamma_5$ is converted into the taste-nonsinglet matrix $(\gamma_5 \otimes \gamma_5^T)$.
The generator $(\gamma_5 \otimes \gamma_5^T)$ defines the U(1)$\times$U(1) residual chiral symmetry, i.e., the exact chiral symmetry of the massless staggered fermion.
The above formulation is based on four tastes, i.e., four flavors.
To implement it in the 2-flavor or (2+1)-flavor dynamical simulation, we need to take square root or fourth root of the Dirac operator.
The subtle problem arises especially in taking the fourth root.
Moreover, since this chiral chemical potential is not taste-singlet, the interpretation of the fourth root is nontrivial.

Since the Wilson Dirac operator (\ref{eqDirac}) is $\gamma_5$-Hermitian, it is straightforward to formulate the domain-wall fermion and the overlap fermion.
This is different from the case of the quark chemical potential, in which the sign function of the non-Hermitian matrix must be introduced.
The domain-wall Dirac operator is
\begin{eqnarray}
D_{\rm dw}(\mu_5) &=& D_{\rm W}(\mu_5) +1 \nonumber\\
&& - \frac{1-\gamma_5}{2} T_{5+} - \frac{1+\gamma_5}{2} T_{5-} \ , 
\end{eqnarray}
where $T_{5\pm}$ is similarly defined to Eqs. (\ref{eqpp}) and (\ref{eqpm}) in the fifth dimension but without the gauge field.
The overlap Dirac operator is
\begin{eqnarray}
D_{\rm ov} (\mu_5) &=& 1 + \gamma_5 \epsilon(\gamma_5 D_{\rm W} (\mu_5)) \ ,
\end{eqnarray}
where $\epsilon(H)$ is the sign function of a Hermitian matrix $H$.
These Dirac operators hold the expected properties, such as the $\gamma_5$-Hermite property and the Ginsparg-Wilson relation.

\end{document}